\documentclass[11pt]{article}

\usepackage[utf8]{inputenc}
\usepackage[T1]{fontenc}
\usepackage{amsmath,amssymb,amsfonts,amsthm}
\usepackage{mathtools}
\usepackage{bm}
\usepackage{booktabs}
\usepackage{array}
\usepackage{multirow}
\usepackage{geometry}
\usepackage[round,authoryear]{natbib}
\usepackage[hidelinks]{hyperref}

\usepackage{graphicx}

\newcommand{\bx}{\mathbf{x}}
\newcommand{\balpha}{\boldsymbol{\alpha}}

\newcommand{\bSigma}{\boldsymbol{\Sigma}}

\newcommand{\SDD}{\mathrm{SDD}}

\newcommand{\E}{\operatorname{E}}
\newcommand{\Var}{\operatorname{Var}}
\newcommand{\Cov}{\operatorname{Cov}}

\newcommand{\sddfigure}[2]{%
  \IfFileExists{#1}{%
    \includegraphics[width=#2]{#1}%
  }{%
    \fbox{%
      \parbox[c][4cm][c]{#2}{%
        \centering
        Figure file not found:\\
        \texttt{\detokenize{#1}}%
      }%
    }%
  }%
}

\title{
\textbf{
On the Spherical-Dirichlet Distribution:
Corrections and Results
}
}

\author{
Jose H. Guardiola\\
Department of Mathematics and Statistics\\
Texas A\&M University-Corpus Christi\\
\texttt{jose.guardiola@tamucc.edu}
}

\date{}

\begin{document}

\maketitle

\begin{abstract}
This article revisits the spherical-Dirichlet
distribution (SDD) introduced by
\citet{Guardiola2020} and clarifies the
correction to its density discussed in
\citet{Guardiola2025}.
The principal issue concerns Equation (3)
of the original publication, which expresses
the density in local Euclidean coordinates
rather than with respect to spherical
surface-area measure.
We derive the corrected intrinsic density
and revisit its moments, covariance matrix,
mode, and limiting properties.
We show that the two density representations
describe the same probability distribution,
preserving the moment formulas and
likelihood-based parameter estimation.
However, the mode and uniformity conditions
depend on the reference measure and require
corresponding clarification.
Reported simulation and text mining estimates are revisited and updated.
The resulting presentation reconciles
the original construction with its
intrinsic spherical properties.
\end{abstract}


\section{Introduction}
\label{sec:intro}

The Spherical-Dirichlet distribution (SDD)
was introduced by \citet{Guardiola2020}
for modeling nonnegative directional data
on the unit hypersphere.
It is obtained by applying a coordinatewise
square-root transformation to a
Dirichlet random vector.

A subsequent correction preprint
\citep{Guardiola2025} addressed the
mathematical formulation of the SDD.
The main issue concerns Equation (3)
of the original publication:
its density is expressed relative to
a coordinate measure rather than
spherical surface-area measure.

This article clarifies that distinction
and presents a consistent formulation
of the density, moments, covariance
structure, mode, and limiting properties.
For completeness, we also revisit
parameter estimation and the original
simulation and text mining applications.

\section{Basic Properties}

\subsection{Density and Normalizing Constant}

Throughout, let $p\geq2$ be an integer. Let
$\boldsymbol{Z}\sim
\operatorname{Dirichlet}(\alpha_1,\ldots,\alpha_p)$,
where $\alpha_i>0$ and
$\alpha_0=\sum_{i=1}^{p}\alpha_i$.
Define $X_i=\sqrt{Z_i}$.
Then $\boldsymbol{X}$ lies on
\[
\mathbb{S}^{p-1}_{+}
=
\left\{
\bx\in\mathbb{R}^{p}:
x_i\geq0,\;
\sum_{i=1}^{p}x_i^2=1
\right\}.
\]

Write $\mathbb S^{p-1}_{+,\circ}=\{\bx\in\mathbb S^{p-1}_+:x_i>0
\text{ for all }i\}$. The random vector lies in this relative interior almost surely;
its topological support is the closed orthant. All density formulas below are
specified on the relative interior. Boundary values can be assigned arbitrarily
without changing the probability law.

The Dirichlet density, with respect to $dz_1\cdots dz_{p-1}$ on
$z_i>0$, $\sum_{i<p}z_i<1$, and $z_p=1-\sum_{i<p}z_i$, is
\[
f_{\mathrm{Dir}}(\boldsymbol{z};\balpha)
=
\frac{\Gamma(\alpha_0)}
{\prod_{i=1}^{p}\Gamma(\alpha_i)}
\prod_{i=1}^{p}z_i^{\alpha_i-1}.
\]

Using $z_i=x_i^2$ for
$i=1,\ldots,p-1$, the coordinate
Jacobian is
\[
\left|
\frac{\partial(z_1,\ldots,z_{p-1})}
{\partial(x_1,\ldots,x_{p-1})}
\right|
=
2^{p-1}\prod_{i=1}^{p-1}x_i.
\]

Consequently, Equation (3) of
\citet{Guardiola2020} can be written as
\begin{equation}
g(\bx;\balpha)
=
\frac{2^{p-1}\Gamma(\alpha_0)}
{\prod_{i=1}^{p}\Gamma(\alpha_i)}
\left(
\prod_{i=1}^{p}x_i^{2\alpha_i-1}
\right)x_p^{-1}.
\label{eq:original-density}
\end{equation}

Here $x_p=(1-\sum_{i<p}x_i^2)^{1/2}$, and this expression is a density
on $D=\{(x_1,\ldots,x_{p-1}):x_i>0,\sum_{i<p}x_i^2<1\}$
with respect to $dx_1\cdots dx_{p-1}$; it is not a density with respect
to $p$-dimensional Lebesgue measure.
However, the spherical surface-area
element satisfies
\[
d\omega_{p-1}(\bx)
=
\frac{dx_1\cdots dx_{p-1}}{x_p}
=
\frac{dz_1\cdots dz_{p-1}}
{2^{p-1}\prod_{i=1}^{p}x_i}.
\]

Indeed, the graph map $F(u)=(u,\sqrt{1-\|u\|^2})$ satisfies
$\det\{(DF)^TDF\}=\det(I+uu^T/x_p^2)=1/x_p^2$.
This proves the surface-area element and its extra factor $1/x_p$.

Thus, the corrected SDD density with
respect to spherical surface-area
measure is
\begin{equation}
f_{\mathrm{SDir}}(\bx;\balpha)
=
\frac{2^{p-1}\Gamma(\alpha_0)}
{\prod_{i=1}^{p}\Gamma(\alpha_i)}
\prod_{i=1}^{p}x_i^{2\alpha_i-1}
,
\qquad
\bx\in\mathbb{S}^{p-1}_{+,\circ}.
\label{eq:SDD}
\end{equation}

In particular,
\[
\int_{\mathbb S^{p-1}_+}\prod_i x_i^{2\alpha_i-1}\,d\omega_{p-1}(x)
=2^{1-p}\frac{\prod_i\Gamma(\alpha_i)}{\Gamma(\alpha_0)},
\]
which verifies the normalizing constant.
We write
$\boldsymbol{X}\sim\SDD(\balpha)$.

\subsection{Moments}

The moments follow directly from
the Dirichlet distribution.
For real $r_i$ satisfying $\alpha_i+r_i/2>0$ for every $i$,
\begin{equation}
\E\left(\prod_{i=1}^{p}X_i^{r_i}\right)
=
\frac{\Gamma(\alpha_0)}
{\Gamma\left(
\alpha_0+\frac12\sum_{i=1}^{p}r_i
\right)}
\prod_{i=1}^{p}
\frac{\Gamma(\alpha_i+r_i/2)}
{\Gamma(\alpha_i)}.
\label{eq:general-moment}
\end{equation}

This follows by substituting $X_i^{r_i}=Z_i^{r_i/2}$ and taking
 the ratio of the Dirichlet normalizing integrals. The stated conditions
 are necessary and sufficient for finiteness of these mixed moments.

Define
\[
m_i=
\frac{\Gamma(\alpha_i+1/2)}
{\Gamma(\alpha_i)},
\qquad
m_0=
\frac{\Gamma(\alpha_0+1/2)}
{\Gamma(\alpha_0)},
\]
and let
$\boldsymbol{m}=(m_1,\ldots,m_p)^T$.
Then
\begin{equation}
\E(X_i)=\frac{m_i}{m_0},
\qquad
\E(X_i^2)=\frac{\alpha_i}{\alpha_0},
\qquad
\E(X_iX_j)=\frac{m_i m_j}{\alpha_0},
\quad i\neq j.
\label{eq:first-moment}
\end{equation}

Consequently,
\begin{equation}
\Var(X_i)
=
\frac{\alpha_i}{\alpha_0}
-\frac{m_i^2}{m_0^2},
\qquad
\Cov(X_i,X_j)
=
m_i m_j
\left(
\frac1{\alpha_0}-\frac1{m_0^2}
\right),
\quad i\neq j.
\label{eq:Var}
\end{equation}

The covariance matrix is
\begin{equation}
\bSigma
=
\frac1{\alpha_0}
\operatorname{diag}
(\alpha_1-m_1^2,\ldots,\alpha_p-m_p^2)
+
\left(
\frac1{\alpha_0}-\frac1{m_0^2}
\right)
\boldsymbol{m}\boldsymbol{m}^T.
\label{eq:covariance}
\end{equation}

These moment expressions remain valid under the corrected density.
They also give $\operatorname{tr}(\bSigma)=1-\|\E\boldsymbol X\|^2$.
All off-diagonal covariances are strictly negative: if
$G\sim\operatorname{Gamma}(\alpha_0,1)$, strict Jensen inequality gives
$m_0=\E\sqrt G<\sqrt{\alpha_0}$, so $1/\alpha_0-1/m_0^2<0$.
Although $\|\boldsymbol X\|=1$, its covariance matrix is positive definite.
For a nonzero vector $a$, $a^T\boldsymbol X$ is not constant on an open
spherical orthant, which has positive density, so $\Var(a^T\boldsymbol X)>0$.
The nonlinear norm constraint does not imply a singular covariance matrix.

\subsection{Mode and Relationship with the Mean}

Maximizing the logarithm of
Equation~\eqref{eq:SDD} subject to
$\sum_{i=1}^{p}x_i^2=1$ gives
\begin{equation}
x_i^{\mathrm{mode}}
=
\sqrt{
\frac{2\alpha_i-1}{2\alpha_0-p}
},
\qquad i=1,\ldots,p,
\label{eq:mode}
\end{equation}
provided $\alpha_i>1/2$ for all $i$.

For completeness, put $b_i=2\alpha_i-1$. On writing $z_i=x_i^2$,
maximization is equivalent to maximizing $\tfrac12\sum_i b_i\log z_i$
on the simplex. If all $b_i>0$, strict concavity proves that
Equation~\eqref{eq:mode} is the unique global mode.
If all $b_i\geq0$ and at least one is positive, the canonical continuous
extension of the density has the unique maximizer
\[
x_i^*=\begin{cases}
\sqrt{b_i/\sum_j b_j},&b_i>0,\\
0,&b_i=0.
\end{cases}
\]
To see this, any mass assigned to a zero-exponent coordinate can be moved
proportionately to the positive-exponent coordinates to increase the product;
strict concavity then applies on that face. For example,
$\balpha=(1/2,1/2,2)$ has its maximum at $(0,0,1)$.
If all $b_i=0$, the density is constant and every point is a maximizer
of its continuous extension.
If some $b_i<0$, let the corresponding coordinate tend to zero while
keeping all remaining coordinates positive. The density diverges, and
there is no finite global density maximum. We therefore describe boundary
singularities rather than asserting a uniquely defined boundary mode.
Boundary maximizers in this discussion refer to the specified continuous
extension, not arbitrary values of an almost-everywhere density.

For the symmetric SDD,
$\alpha_i=\alpha>1/2$,
\begin{equation}
\boldsymbol{x}^{\mathrm{mode}}
=
\frac1{\sqrt p}\boldsymbol{1}.
\label{eq:symmetric-mode}
\end{equation}

The mean vector is
\begin{equation}
\E(\boldsymbol{X})
=
\frac{
\Gamma(\alpha+1/2)\Gamma(p\alpha)
}{
\Gamma(\alpha)\Gamma(p\alpha+1/2)
}
\boldsymbol{1}.
\label{eq:symmetric-mean}
\end{equation}

The mean and mode have the same
direction, although the mean has
norm less than one for finite
$\alpha$.
Using the asymptotic gamma ratio,
\[
\frac{\Gamma(t+1/2)}{\Gamma(t)}
\sim t^{1/2},
\qquad t\to\infty,
\]
we obtain
\[
\E(\boldsymbol{X})
\longrightarrow
\frac1{\sqrt p}\boldsymbol{1}
\qquad\text{as }\alpha\to\infty.
\]

The mode is defined with respect to spherical surface-area
measure. Under this formulation, all coordinates and their
corresponding parameters enter symmetrically. The unique
interior mode is
\[
x_i^{\mathrm{mode}}
=
\sqrt{\frac{2\alpha_i-1}{2\alpha_0-p}},
\qquad i=1,\ldots,p,
\]
provided that $\alpha_i>1/2$ for every $i$, where
$\alpha_0=\sum_{j=1}^{p}\alpha_j$.
The apparent asymmetry in the original coordinate density
arises solely from the choice of local coordinates; it is
absent from the intrinsic surface-area density.

\section{Relationships with Other Distributions}

\subsection{Limiting and Uniform Cases}

For the symmetric SDD,
$\alpha_i=\alpha$,
\begin{equation}
f_{\mathrm{SDir}}(\bx;\alpha)
=
\frac{2^{p-1}\Gamma(p\alpha)}
{\Gamma(\alpha)^p}
\prod_{i=1}^{p}x_i^{2\alpha-1}.
\end{equation}

Since
$X_i^2\sim
\operatorname{Beta}(\alpha,(p-1)\alpha)$,
\[
\E(X_i^2)=\frac1p,
\qquad
\Var(X_i^2)
=
\frac{p-1}{p^2(p\alpha+1)}.
\]
Consequently,
\begin{equation}
\boldsymbol{X}
\xrightarrow{\mathbb{P}}
\frac1{\sqrt p}\boldsymbol{1}
\qquad\text{as }\alpha\to\infty.
\label{eq:symmetric-limit}
\end{equation}

When $\alpha_i=1/2$ for all $i$,
Equation~\eqref{eq:SDD} reduces to
\begin{equation}
f_{\mathrm{SDir}}(\bx)
=
\frac{2^{p-1}\Gamma(p/2)}{\pi^{p/2}},
\qquad
\bx\in\mathbb{S}^{p-1}_{+}.
\label{eq:uniform}
\end{equation}

The area of the positive orthant is
$\omega_{p-1}(\mathbb S^{p-1}_+)=2^{1-p}\pi^{p/2}/\Gamma(p/2)$,
so the displayed density is exactly its reciprocal. For $p\geq2$, all
$\alpha_i=1/2$ is also necessary for a constant surface density:
$\sum_i(\alpha_i-1/2)\log z_i$ can be constant throughout the simplex
only when all its coefficients vanish.

This differs from the coordinate
uniformity condition in the original
article, which used
$\alpha_i=1/2$ for $i<p$ and
$\alpha_p=1$.

\subsection{Comparison with Directional Distributions}

The von Mises--Fisher distribution has
density
\[
f_{\mathrm{vMF}}(\bx;\boldsymbol{\mu},\kappa)
=
c_p(\kappa)
\exp(\kappa\boldsymbol{\mu}^T\bx),
\]
where $\bx\in\mathbb S^{p-1}$, $c_p(\kappa)$ is the normalizing constant
with respect to full-sphere surface area, $\|\boldsymbol{\mu}\|=1$, and
$\kappa\geq0$ \citep{mardiajupp}.

More generally, the Fisher--Bingham
family has density proportional to
\[
f_{\mathrm{FB}}(\bx)
\propto
\exp\left(
\boldsymbol{b}^T\bx+\bx^T A\bx
\right),
\]
where $A$ is symmetric
\citep{Kent,mardiajupp}.

Unlike their usual full-sphere
formulations, the SDD is supported
only on the positive orthant.
Its parameters jointly determine
shape and concentration. Truncating the other families to this orthant may yield analytically intractable normalizing constants, requiring numerical approximation.

\section{Inference for the
Spherical-Dirichlet Distribution}

Let
$\boldsymbol{X}_1,\ldots,\boldsymbol{X}_n$
be independent observations from
$\SDD(\balpha)$, with every coordinate strictly positive.
Exact observed zeros require a separate observation model or an explicitly
justified preprocessing rule; the log-likelihood below does not apply to them directly.

\subsection{Method of Moments}

Define
\[
\overline{X}_j
=
\frac1n\sum_{k=1}^{n}X_{kj},
\qquad
\overline{X_j^2}
=
\frac1n\sum_{k=1}^{n}X_{kj}^2.
\]

Since
$\sum_{j=1}^{p}\overline{X_j^2}=1$,
only $p-1$ second-moment equations
are independent.

Following the original approach,
we use
\begin{equation}
\frac{\Gamma(\alpha_1+1/2)}
{\Gamma(\alpha_1)}
\frac{\Gamma(\alpha_0)}
{\Gamma(\alpha_0+1/2)}
=
\overline{X}_1
\label{eq:06}
\end{equation}
and
\begin{equation}
\frac{\alpha_j}{\alpha_0}
=
\overline{X_j^2},
\qquad j=2,\ldots,p.
\label{eq:07}
\end{equation}

Equivalently,
$\alpha_j=\alpha_0\overline{X_j^2}$
for all $j$.
Substituting into
Equation~\eqref{eq:06}
reduces estimation to
a scalar nonlinear equation
for $\alpha_0$.
The MOM estimates are then
\[
\widehat{\alpha}_j^{\mathrm{MOM}}
=
\widehat{\alpha}_0\,
\overline{X_j^2}.
\]

\paragraph{Existence, uniqueness, and computation.}
Put $q=\overline{X_1^2}\in(0,1)$ and $a=\overline X_1$. The scalar equation is
\[
 h_q(t)=\frac{\Gamma(tq+1/2)\Gamma(t)}{\Gamma(tq)\Gamma(t+1/2)}=a,
 \qquad t>0.
\]
It has a unique finite solution if and only if $q<a<\sqrt q$.
Indeed, $h_q(0+)=q$ and $h_q(\infty)=\sqrt q$.
To prove strict monotonicity, put $d(s)=\psi(s+1/2)-\psi(s)$. Its integral
representation yields
\[
 s d(s)=\int_0^\infty\frac{e^{-v}}{1+e^{-v/(2s)}}\,dv,
\]
which is strictly decreasing in $s$. Hence
$\{\log h_q(t)\}'=q d(tq)-d(t)>0$.
For interior observations, $q<a$ since $X_{k1}^2<X_{k1}$;
$a\leq\sqrt q$ follows from Cauchy--Schwarz, with strict inequality
exactly when the first coordinate is not constant across the sample.
A constant first coordinate therefore gives no finite MOM solution,
even if other coordinates vary. Another nonconstant coordinate can be
chosen in that case. Different choices generally produce different estimates.
A bracketed scalar root finder with log-gamma evaluation avoids unrestricted
multivariate optimization and guarantees convergence once a finite root is bracketed.

\subsection{Maximum Likelihood Estimation}

The log-likelihood is
\begin{align}
\ell(\balpha)
&=
n(p-1)\log2
+n\log\Gamma(\alpha_0)
\nonumber\\
&\quad
-n\sum_{j=1}^{p}\log\Gamma(\alpha_j)
\nonumber\\
&\quad+
\sum_{j=1}^{p}
(2\alpha_j-1)
\sum_{k=1}^{n}\log X_{kj}.
\label{eq:loglik}
\end{align}

The score equations are
\begin{equation}
\psi(\alpha_j)-\psi(\alpha_0)
=
\frac2n\sum_{k=1}^{n}\log X_{kj},
\qquad j=1,\ldots,p,
\label{eq:mle-score}
\end{equation}
where $\psi$ is the digamma function.

For $n\geq1$, the negative log-likelihood is strictly convex in
$\balpha\in(0,\infty)^p$. Its Hessian is
\[
 H=n\left[\operatorname{diag}\{\psi_1(\alpha_j)\}
       -\psi_1(\alpha_0)\boldsymbol 1\boldsymbol 1^T\right],
\]
where $\psi_1$ is the trigamma function. This is
$n\operatorname{Cov}(\log\boldsymbol Z)$ under the model, and is positive
definite: a nonzero linear combination of $\log Z_j$ cannot be constant
on the open simplex. Thus any finite stationary point is the unique global
maximum of the log-likelihood. Strict convexity alone does not establish
existence of such a point. In particular, if all observations equal an
interior vector $x$, the sequence $\alpha_j=t x_j^2$ has
$\ell(\balpha)=n(p-1)\log(t)/2+O(1)\to\infty$, so no finite MLE exists.
This includes $n=1$.
Numerical minimization can be
performed using L-BFGS-B,
subject to $\alpha_j\geq\varepsilon>0$. A fixed lower bound gives a
constrained estimate, which agrees with the unrestricted MLE only when the
latter lies in the feasible set. 
Strict convexity here is in $\balpha$, and is not automatically preserved
by a log-parameter reparameterization.

The gradient is
\begin{equation}
\frac{\partial(-\ell)}
{\partial\alpha_j}
=
n\left[
\psi(\alpha_j)-\psi(\alpha_0)
\right]
-
2\sum_{k=1}^{n}\log X_{kj}.
\label{eq:gradient}
\end{equation}

Importantly,
Equation~\eqref{eq:original-density}
differs from the corrected density
by a factor independent of
$\balpha$.
Therefore, the correction does not
change the likelihood score equations
or the maximum likelihood estimator.

\section{Applications to Data}

We revisit the simulation and
text mining applications from
\citet{Guardiola2020}. The following numerical results
represent updated analyses.

\subsection{Simulation Example}

Four parameter configurations were
considered for a three-dimensional SDD:
\[
(2,2,2),\quad
(5,15,2),\quad
(0.5,0.5,2),\quad
(2,2,10).
\]

Each simulation generated 10,000
observations using
$X_j=\sqrt{Z_j}$, where
$\boldsymbol{Z}\sim
\operatorname{Dirichlet}(\balpha)$.

\begin{figure}[htbp]
\centering
\begin{minipage}{0.48\textwidth}
\centering
\includegraphics[width=\linewidth]{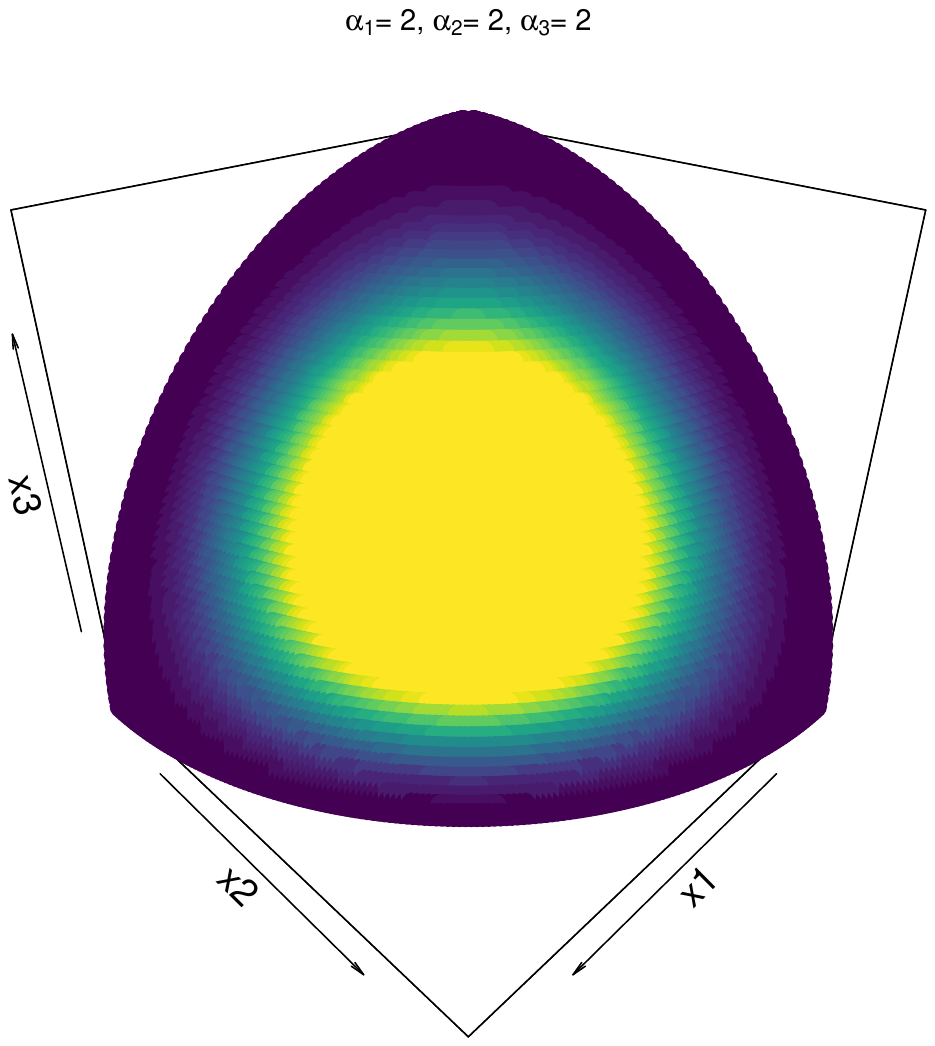}

\small (a) $\balpha=(2,2,2)$
\end{minipage}
\hfill
\begin{minipage}{0.48\textwidth}
\centering
\includegraphics[width=\linewidth]{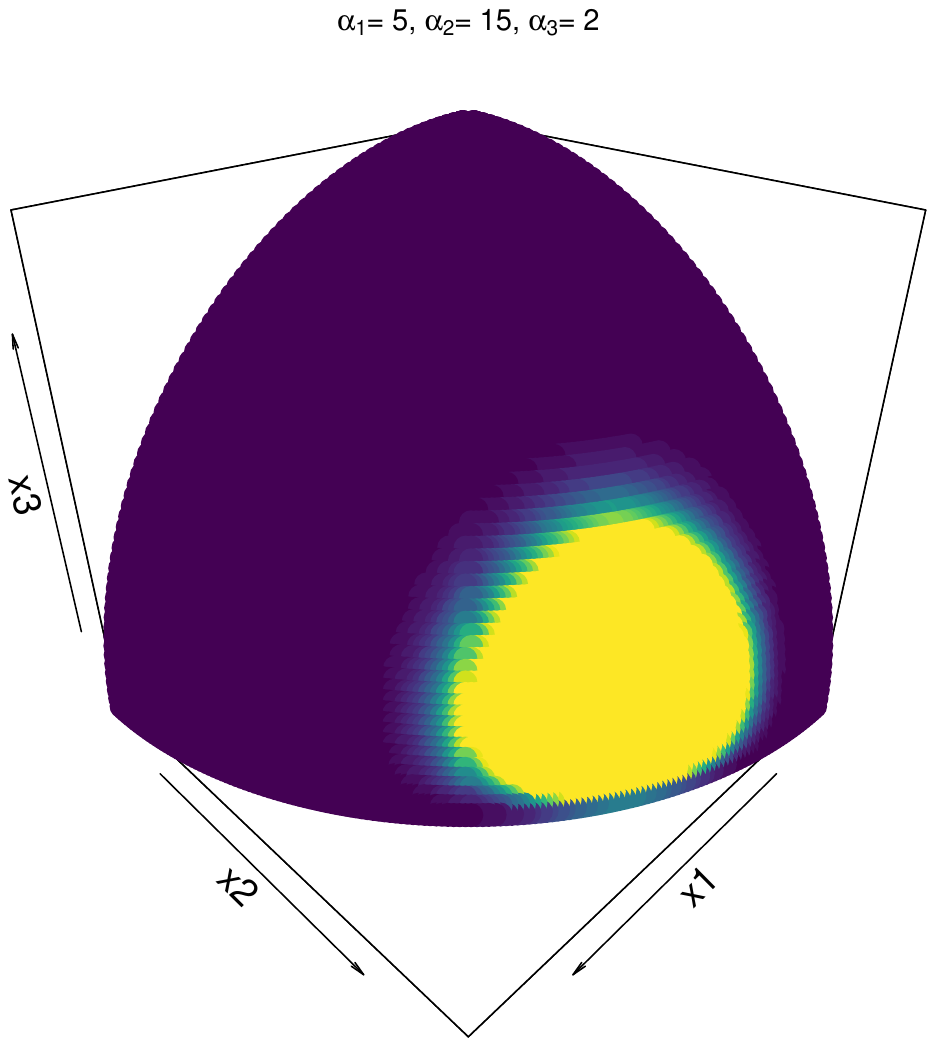}

\small (b) $\balpha=(5,15,2)$
\end{minipage}

\vspace{0.3cm}

\begin{minipage}{0.48\textwidth}
\centering
\includegraphics[width=\linewidth]{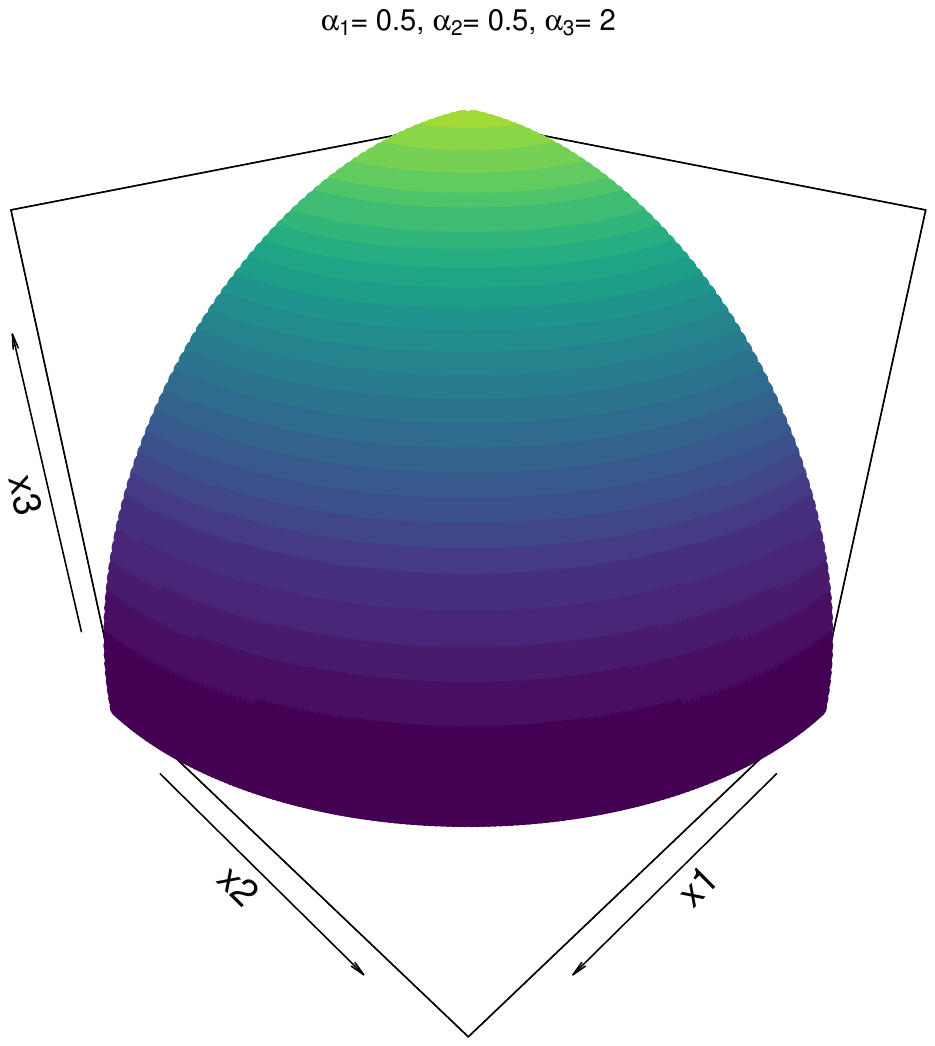}

\small (c) $\balpha=(0.5,0.5,2)$
\end{minipage}
\hfill
\begin{minipage}{0.48\textwidth}
\centering
\includegraphics[width=\linewidth]{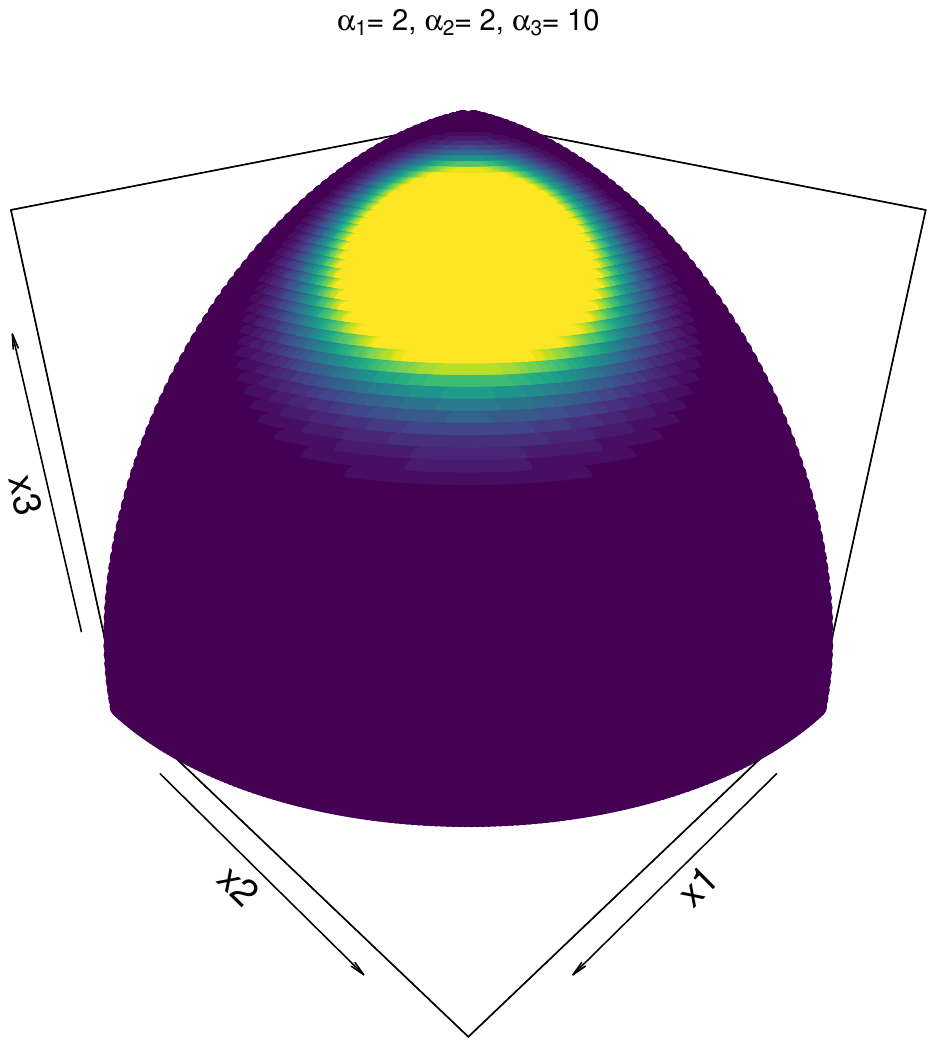}

\small (d) $\balpha=(2,2,10)$
\end{minipage}

\caption{Spherical-Dirichlet distributions
under four concentration parameter
configurations.}
\label{fig:simulation}
\end{figure}

The corresponding distributions
are illustrated in
Figure~\ref{fig:simulation}.

The estimation error was measured by
the relative Euclidean norm
$\operatorname{Error}(\%)
=
100\|\widehat{\balpha}-\balpha\|_2/
\|\balpha\|_2$.

Table~\ref{tab:simulation}
reports the parameter estimates,
iteration counts, and relative errors.

\begin{table}[htbp]
\centering
\caption{Simulation estimates; errors recomputed from displayed values.}
\label{tab:simulation}
\small
\setlength{\tabcolsep}{5pt}
\begin{tabular}{lccccc}
\toprule
Method & Iter. &
$\hat{\alpha}_1$ &
$\hat{\alpha}_2$ &
$\hat{\alpha}_3$ &
Error (\%)\\
\midrule
\multicolumn{6}{c}{$\balpha=(2,2,2)$}\\
MOM & 176 & 2.0557 & 2.0983 & 2.0764 & 3.94\\
MLE & 8 & 2.0538 & 2.0433 & 2.0426 & 2.34\\
\midrule
\multicolumn{6}{c}{$\balpha=(5,15,2)$}\\
MOM & 589 & 5.0351 & 14.7148 & 1.9684 & 1.81\\
MLE & 19 & 5.0584 & 15.2103 & 2.0158 & 1.37\\
\midrule
\multicolumn{6}{c}{$\balpha=(0.5,0.5,2)$}\\
MOM & 28 & 0.4964 & 0.4639 & 1.9212 & 4.09\\
MLE & 13 & 0.4997 & 0.5004 & 2.0091 & 0.43\\
\midrule
\multicolumn{6}{c}{$\balpha=(2,2,10)$}\\
MOM & 385 & 1.9349 & 2.0153 & 10.1145 & 1.28\\
MLE & 14 & 2.0190 & 2.0159 & 10.1229 & 1.21\\
\bottomrule
\end{tabular}
\end{table}

The displayed estimates have small relative errors. In these four reported
samples, the MLE errors are smaller than the MOM errors. This does not
establish uniformly better finite-sample performance. The MOM estimates
and iteration counts coincide with those in \citet{Guardiola2020}, whereas
the MLE entries differ slightly.
Iteration counts from different algorithms are not directly comparable measures of computational cost.

\subsection{Text Mining Example}

The text mining application used
160 documents from the automobile
discussion category of the CMU
text collection \citep{CMU}.

After preprocessing, nine terms
were retained:
\textit{auto, write, articl, engin,
don, good, time, drive}, and
\textit{road}.

Term frequencies were transformed
using
$x_{\mathrm{transf}}=\log(1.10+x)$
and normalized to unit Euclidean
length. The resulting observations
lie in the positive orthant of
the eight-dimensional unit sphere $\mathbb S^8$ embedded in $\mathbb R^9$.

Table~\ref{tab:text-results}
reports the MOM and MLE estimates.

\begin{table}[htbp]
\centering
\caption{Text mining parameter estimates (nine components).}
\label{tab:text-results}
\begin{tabular}{lcc}
\toprule
Parameter & MOM & MLE\\
\midrule
$\alpha_1$ & 1.2817 & 1.2743\\
$\alpha_2$ & 0.5575 & 0.5543\\
$\alpha_3$ & 0.4953 & 0.4924\\
$\alpha_4$ & 0.3135 & 0.3138\\
$\alpha_5$ & 0.3624 & 0.3630\\
$\alpha_6$ & 0.3174 & 0.3177\\
$\alpha_7$ & 0.3352 & 0.3356\\
$\alpha_8$ & 0.3100 & 0.3102\\
$\alpha_9$ & 0.2722 & 0.2721\\
\bottomrule
\end{tabular}
\end{table}

The two methods produced
closely agreeing estimates.
MOM required 276 iterations,
compared with 18 for MLE.

The transformation makes every coordinate positive, including zero counts.
It defines a model for the transformed directions, not directly for raw counts.
The agreement of two estimators does not necessarily establish goodness of fit.
The nine-component specification is an intentional correction to the
ten-component description in the 2020 article, as confirmed by the author.
The nine terms listed above define the corrected variables.

\section{Conclusions}

This article clarifies the
density formulation of the
Spherical-Dirichlet distribution
introduced by \citet{Guardiola2020}
and reconciles it with the
subsequent correction
\citep{Guardiola2025}.

The principal distinction is
between the original coordinate
density and the intrinsic density
with respect to spherical
surface-area measure.

The correction preserves the moment formulas,
covariance structure, and likelihood-based
parameter estimation procedures. However,
the mode and uniformity conditions must be
interpreted relative to the reference measure.
In particular, the corrected density yields the intrinsic
spherical mode and includes the uniform distribution with
respect to spherical surface-area measure when
$\alpha_i=1/2,\ \forall i\in\{1,\ldots,p\}$.

In the four simulation examples considered, both methods
produced estimates close to the generating parameters,
with MLE yielding smaller relative estimation errors
than MOM in each example. In the text mining example, MOM and MLE produced closely agreeing parameter estimates for the nine-component transformed document vectors, illustrating the application
of the SDD to strictly positive directional representations
of text data.

In summary, the SDD remains a tractable competitive model for directional data with strictly positive components on the unit hypersphere, with an explicit normalizing constant.

\subsection*{Acknowledgements}

The author thanks Eduardo
Garc\'ia-Portugu\'es, Department
of Statistics, Universidad Carlos III
de Madrid, Spain, for his
valuable assistance.


\bibliographystyle{apalike}
\bibliography{HypersphereBib}

@article{Guardiola2020,
   title={The spherical-{D}irichlet distribution},
   volume={7},
   ISSN={2195-5832},
   url={http://dx.doi.org/10.1186/s40488-020-00106-9},
   DOI={10.1186/s40488-020-00106-9},
   number={1},
   journal={Journal of Statistical Distributions and Applications},
   publisher={Springer Science and Business Media LLC},
   author={Guardiola, Jose},
   year={2020},
   month=sep }

@misc{Guardiola2025,
  author        = {Guardiola, Jose},
  title         = {On the spherical {D}irichlet distribution: corrections and Results},
  year          = {2025},
  eprint        = {2506.04441},
  archivePrefix = {arXiv},
  primaryClass  = {stat.ME},
  url           = {https://arxiv.org/abs/2506.04441}
}

@article{Kent,
  author = {Kent, John T.},
  title = {The {Fisher--Bingham} Distribution on the Sphere},
  journal = {Journal of the Royal Statistical Society:
             Series B (Methodological)},
  volume = {44},
  number = {1},
  pages = {71--80},
  year = {1982},
  doi = {10.1111/j.2517-6161.1982.tb01189.x}
}

@book{mardiajupp,
  author = {Mardia, K. V. and Jupp, P. E.},
  title = {Directional Statistics},
  publisher = {Wiley},
  address = {Chichester, England},
  series = {Wiley Series in Probability and Statistics},
  year = {2000},
  isbn = {0471953334}
}

@misc{CMU,
  author = {Lang, Ken},
  title = {{CMU} Text Learning Group Data Archives},
  year = {2019},
  url = {https://www.cs.cmu.edu/afs/cs/project/theo-20/www/data/news20.html},
  note = {Accessed September 1, 2019}
}
\end{document}